\documentclass[prd,aps,floats,twocolumn,showpacs]{revtex4}
\usepackage{epsfig}

\newcommand{\be}{\begin{equation}}
\newcommand{\ee}{\end{equation}}
\newcommand{\bea}{\begin{eqnarray}}
\newcommand{\eea}{\end{eqnarray}}

\begin{document}

\title{Observational Constraints on the Completeness of Space near
Astrophysical Objects}
\author{Timothy Clifton$^{1}$\footnote{tclifton@astro.ox.ac.uk} 
and 
John D. Barrow$^{2}$\footnote{j.d.barrow@damtp.cam.ac.uk}}
\affiliation{$^1$Department of Astrophysics, University of Oxford, UK.\\
$^2$DAMTP, University of Cambridge, UK. }

\begin{abstract}
We consider the observational effects of a deficit angle, $w$, in the
topology of the solar system and in the `double pulsar' system PSR
J0737-3039A/B. Using observations of the perihelion precession of Mercury,
and the gravitational deflection of light due to the Sun, we constrain the magnitude of such
a deficit angle in the solar system to be $2\pi (1-w)$, with $0\leq
(1-w)<10^{-9}$ at $95\%$ confidence. We calculate the effects of a deficit
angle on the periastron advance, geodetic precession rate and inclination angle of the double pulsar
system and use the observational data to obtain the constraint $0\leq
(1-w)<2.4\times 10^{-8}$ at $95\%$ confidence. Although this result is weaker
than the solar system bound, it is in a very different physical environment,
where accumulating data is likely to lead to tighter constraints in
the future.
\end{abstract}

\date{March 10, 2010}
\maketitle

\section{Introduction}

Metric-based gravity theories, like Einstein's and its close relatives, are
routinely tested using observations of astrophysical systems where
gravitational fields are strong and non-Newtonian. Following Eddington's
introduction of a parameterised form of the metric in 1922 \cite{edd}, and
Nordvedt's extension to more general configurations by including further
parameters \cite{nord}, a sophisticated frame-work has been devised \cite%
{Will}, and a large number of careful observations made, in order to place
stringent constraints on the geometry of space-time. However, most of these
studies have assumed that the astrophysical systems they consider should
have a trivial space-time topology. Under such an assumption, it is possible
to make strong statements about the magnitude of any deviations from the
predictions of general relativity. Here we take a different approach: We
assume a geometry that is locally isometric to the predictions of general
relativity, but which permits non-trivial global topologies. The
astrophysical observations can then be used to constrain the topology of the
systems in question. Of particular interest for this study is the recently
discovered `double pulsar' system, PSR J0737-3039A/B \cite{dp,dp2}.

Non-trivial topologies can exist in a number of astrophysically interesting
situations, including magnetic monopoles, cosmic strings, domain walls and
textures \cite{shellard}, all of which can arise in phase transitions in the
early universe \cite{kibble}. A tell-tale sign of such topological defects
is the existence of a `deficit angle', whereby a wedge of space-time appears
to have been removed, and the surfaces that remain have been identified. A
simple example of this feature, first found by Marder in 1959 \cite{marder},
is the line-element 
\begin{equation}
ds^{2}=-\alpha dt^{2}+\frac{dr^{2}}{\alpha }+r^{2}d\theta
^{2}+w^{2}r^{2}\sin ^{2}\theta d\phi ^{2},  \label{schw1}
\end{equation}%
where $\alpha =1-2GM/r$ and $w\in ( 0,1\rbrack$ is a constant. This is well
known as the exact solution of a point-like mass on an infinitely thin
cosmic string \cite{marder, vilenkin}. It is clearly locally isometric to the
Schwarzschild solution, but if we allow $\phi $ to run from $0$ to $2\pi $,
then we see it has a different topology.

The line-element (\ref{schw1}) may initially appear a trivial manipulation
of the Schwarzschild solution. The effect of introducing such a topological
defect, however, has non-trivial consequences for the geodesic equations.
For the metric above, these equations have recently been solved completely
by Hackmann et al. \cite{hack}. The missing wedge of space-time causes new
behaviour that is not present in the Schwarzschild solution, such as a
precession of the angular momentum vector about the axis of the string \cite%
{masar}. By using observational constraints on such phenomena we can
therefore constrain the amplitude of any possible deficit angle, and hence
place constraints on the topologies of astrophysical systems.

In Section \ref{ss} we summarise how deficit angles affect the geodesic
equations, and observations of gravitational phenomena in the solar system.
in Section \ref{pulsar}, we proceed to consider the double pulsar. This
over-constrained system has a large number of well measured observables, and
provides an excellent laboratory to perform the type of test we are
considering. In Section \ref{conc} we give our conclusions.

\section{Solar system constraints}

\label{ss}

If we choose coordinates so that the orbit we consider is in a plane of
constant elevation ($\theta =\pi/2$), then the effect of a
deficit angle on the equations of motion enters only through the azimuthal
coordinate, $\phi $. To include such a defect, we can then write the range
of $\phi $ as 
\begin{equation}
0\leq \phi \leq 2\pi -\Delta \varphi
\end{equation}%
where $\Delta \varphi \in \lbrack 0,2\pi )$ is the deficit angle. Alternatively, we
can change variables so that $\Delta \varphi =2\pi (1-w)$. Now, $\phi $
covers the range 
\begin{equation}
0\leq \phi \leq 2\pi w,
\end{equation}%
which is equivalent to a simple coordinate redefinition 
\begin{equation}
\phi \rightarrow w\phi .  \label{redef}
\end{equation}

\subsection{Time-like orbits}

Using the redefinition (\ref{redef}), the geodesic equations become 
\begin{equation}
u^{\prime \prime }+w^{2}u=\frac{GM}{L^{2}}+3GMw^{2}u^{2}  \label{tl}
\end{equation}%
where $u=1/r$, and where primes denote differentiation with respect to $\phi $.
The angular momentum constant, $L$, is given by 
\begin{equation}
L=r^{2}\sin (\theta _{0})\frac{d\phi }{d\lambda },  \label{am}
\end{equation}%
where $\lambda $ parameterises distance along the curve. The solution to (%
\ref{tl}) is then given by 
\begin{equation}
u=w^{-2}u_{GR}(w\phi ),
\end{equation}%
where $u_{GR}(\phi )$ is the usual general relativistic solution that
occurs when $w=1$. Therefore, we have 
\begin{equation}
u=\frac{1}{r}=\frac{GM}{w^{2}L^{2}}\left[ 1+e\cos \{w(\phi -\phi _{0}-\Delta
\phi _{0})\}\right]  \label{tlprec}
\end{equation}%
where $e$ is the eccentricity of the orbit, $\phi _{0}$ specifies the phase,
and $\Delta \phi _{0}$ gives the correction due to relativistic perihelion
precession.

It can now be seen from (\ref{tlprec}) that the perihelion shift per orbital
revolution is 
\begin{equation}
w\Delta \phi =\frac{6\pi G^{2}M^{2}}{w^{2}L^{2}}\ +2\pi (1-w).
\end{equation}%
If we now write 
\begin{equation}
w=1-\delta w,  \label{w}
\end{equation}%
where $\delta w\in \lbrack 0,1)$ is small, then to first order we should
expect the precession to be 
\begin{equation}
\Delta \phi \simeq \Delta \phi _{GR}+2\pi \delta w,
\end{equation}%
where $\Delta \phi _{GR}=\{\Delta \phi \}_{w=1}$ is just the usual general
relativistic prediction with no defect. The precision of the agreement with
the current data therefore gives a bound on $\delta w$, and hence on $w$.

In order to gain observations of $\Delta \phi $ from the precession of the
orbit of Mercury,
with respect to the vernal equinox of the
Sun, it is necessary to take into account the precession of the equinoxes on
the coordinate system (about 5025$^{\prime\prime}$ per century), the perturbing effects of
the other planets (about 531$^{\prime\prime}$ per century) and the effect of the quadrupole
moment of the Sun (about 0.025$^{\prime\prime}$ per century), on the perihelion precession
of the objects that orbit it. These challenges can be addressed in a variety
of different ways, and the result is authors claiming slightly different
observational bounds on the residual $\Delta \phi $. Rather than favoring
any particular method here, we prefer to quote a number of different precise
calculations, obtained by various authors. These are displayed in Table %
\ref{prec}. For further details the reader is referred to the original
papers, and references therein. For an overview of the issues involved the
reader is referred to \cite{precpub}.

\begin{table}[tbh]
\begin{center}
\begin{tabular}{|l|c|c|}
\hline
\qquad Source & $\Delta \phi-\Delta \phi_{GR}$ & $10^{10} \times \delta w$
\\ 
& (arcsec/century) & ($2\sigma $ upper bound) \\ \hline
Anderson \textit{et al.} \cite{91} & $-0.04 \pm 0.20$ & 6.7 \\ 
Anderson \textit{et al.} \cite{92} & $+0.15 \pm 0.14$ & 8.0 \\ 
Krasinsky \textit{et al.} \cite{93a} &  &  \\ 
\qquad EPM$1988$ & $+0.004 \pm 0.061$ & 2.3 \\ 
\qquad DE$200$ & $-0.003 \pm 0.061$ & 2.2 \\ 
Pitjeva \cite{93b} &  &  \\ 
\qquad EPM$1988$ & $-0.017 \pm 0.052$ & 1.6 \\ 
\qquad DE$200$ & $-0.011 \pm 0.052$ & 1.7 \\ \hline
\end{tabular}%
\end{center}
\caption{The value of the perihelion precession of Mercury obtained from
observations by various authors, and the resulting $2\protect\sigma $
upper bound on the permissible values of $\protect\delta w$. The acronyms
EPM$1988$ and DE$200$ refer to different numerical ephemerides, which are reviewed in 
\protect\cite{pit01}. We take the sidereal period of Mercury to be $0.24$
years \protect\cite{allen}.}
\label{prec}
\end{table}

It can be seen from Table \ref{prec} that placing a constraint on $\delta w$
is not a clear-cut matter, and depends both on the data used, and how it is
treated. However, all of the results in Table \ref{prec} are consistent with
the conservative statement that the $2\sigma $ bound on $\delta w$ from
observations of the perihelion precession of Mercury is within the range 
\begin{equation}
0\leq \delta w<10^{-9}.  \label{tlbound}
\end{equation}

\subsection{Null orbits}

The geodesic equation for light rays is transformed in a similar way, but
there is now no $L$. Under the redefinition (\ref{redef}), we therefore have
simply 
\begin{equation}
u^{\prime \prime }+w^{2}u=3GMw^{2}u^{2},
\end{equation}%
which has the solution 
\begin{equation}
u=u_{GR}(w\phi ),
\end{equation}
where, again, $u_{GR}(\phi )$ is the usual general relativistic solution
with $w=1$. Taking this, together with $\phi =\pi +\delta \phi $, we obtain 
\begin{equation}
u=\frac{1}{r}=\frac{4GM}{R^{2}}-\frac{w\delta \phi }{R}+\frac{\pi }{R}(1-w),
\end{equation}%
where $R$ is the impact parameter. The total deflection is then 
\begin{equation}
w\delta \phi =\frac{4GM}{R}+\pi (1-w)= \frac{4GM}{\ R}+\pi \delta w,
\end{equation}%
and the deflection caused by the defect is 
\begin{equation}
\delta \phi \simeq \delta \phi _{GR}+\pi \delta w,  \label{def}
\end{equation}%
where $\delta \phi _{GR}=\{\delta \phi \}_{w=1}$ is the usual general
relativistic prediction. The bound on $\delta w$, and hence $w$, can now be
obtained by comparing (\ref{def}) with observed deflections.

The best results available on light bending in gravitational fields
are those of Shapiro, Davis, Lebach and Gregory \cite{sdlg}. These authors use almost
2 million observations of 541 radio sources by 87 Very-Long-Baseline
Interferometry (VLBI) sites to calculate the deflection caused by the
gravitational field of the Sun. Their result is 
\begin{equation}
\delta \phi =\left( 0.99992\pm 0.00023\right) \delta \phi _{GR},
\end{equation}%
which gives us a $2\sigma $ bound on $\delta w$ of 
\begin{equation}
0\leq \delta w<9.9\times 10^{-10},  \label{llbound}
\end{equation}%
where we have taken the general relativistic prediction of light bending for
an object whose light ray grazes the Sun's limb to be $1.75^{\prime \prime}$.
This is almost exactly the same bound as was achieved for the time-like case 
\footnote{%
Note that these bounds are weaker than those of \cite{hack}, who appear to
have erroneously used constraints on post-Newtonian parameters derived from
the Nordvedt effect and the Shapiro effect to constrain perihelion
precession and light bending.}.

\section{Pulsar constraints}

\label{pulsar}

The recent discovery of the `double pulsar' system \cite{dp,dp2}, PSR J$%
0737$-$3039$A/B, provides new opportunities for testing ideas in relativistic
gravity \cite{1,2}. These binary neutron stars are known to orbit each other
with a period of $~$2.45 hours, with much higher velocities and
accelerations than those found in other binary pulsar systems. By good
fortune, this system is also relatively near to the Sun, and oriented so
that we observe it nearly edge-on, at an inclination angle of about $%
89^{\circ }$. Most importantly, however, and unique to PSR J$0737$-$3039$A/B, is the
feature that both neutron stars are detectable as radio pulsars, with
periods of $22$ms and $2.7$s, respectively for PSR J$0737$-$3039$A and PSR J$%
0737$-$3039$B. These properties make the double pulsar an excellent tool for
constraining the type of deviant gravitational phenomena we are considering
in this article.

To proceed further, we model the space-time geometry as 
\begin{equation}
ds^{2}=-(1-2U)dt^{2}+(1+2U)\tilde{\delta}_{ij}dx^{i}dx^{j}  \label{geo}
\end{equation}%
where $\tilde{\delta}_{ij}$ specifies the static geometry of the 3-space. We
can now assign orders of smallness in the usual way, so that the Newtonian
potential $U$ is $O(2)$ small, time derivatives add an $O(1)$ of smallness,
and the 3-velocity $v^{i}$ is $O(1)$ small.


We will now consider relativistic effects that are good candidates for
constraining the existence of any deficit angle in the double pulsar, before
comparing our predictions with the observational data.

\subsection{Spin Precession}

Geodetic precession of a body's spin vector about its orbital angular
momentum vector is a well known prediction of relativistic gravity, and has
already been studied in some binary pulsar systems \cite{2,8,9,wt,10,cw}.
However, while it took over a decade for geodetic precession to be observed
in the Hulse-Taylor pulsar, PSR B$1913$+$16$ \cite{9}, it has already been
reported in the double pulsar \cite{2}. We expect this effect to be
particularly sensitive to the existence of a deficit angle, as it involves
integration over an orbit. We shall therefore calculate its
influence below, before proceeding to infer constraints from observational
data later in the section.

First, we want the pulsar spin vector, $Y^{\mu }$, to be orthogonal to the
world-line of a particle, $u^{\mu }$, and to be parallel propagated along
that curve, so that 
\begin{equation}
\frac{dY_{\mu }}{d\tau }=g^{\lambda \sigma }\Gamma _{\sigma \mu \nu
}Y_{\lambda }\frac{dx^{\nu }}{d\tau }=\Gamma _{\sigma \mu \nu }Y^{\sigma }%
\frac{dx^{\nu }}{d\tau },
\end{equation}%
where $\Gamma _{\mu \nu \sigma }=g_{\mu \rho }{\Gamma ^{\rho }}_{\nu \sigma
},$ and $\tau $ is proper time along the particle's world-line. The
orthogonality condition then gives 
\begin{equation}
Y_{0}=-v^{i}Y_{i}+O(3),
\end{equation}%
where $v^{i}$ is the 3-velocity of the particle. Multiplying through by $%
d\tau /dt$ we find that the spatial component of $Y^{\mu }$ obeys 
\begin{eqnarray}
\frac{dY_{i}}{dt} &=&\Gamma _{0i0}v^{j}Y_{j}+\Gamma _{ji0}Y^{j}+\Gamma
_{kij}v^{j}Y^{k}+O(5) \\
&=&\tilde{\delta}_{ij}U_{,0}Y^{j}+(2\tilde{\delta}_{kj}U_{,i}+\tilde{\delta}%
_{ik}U_{,j}-\tilde{\delta}_{ij}U_{,k})v^{j}Y^{k}  \nonumber \\
&&+(1+2U)\tilde{\Gamma}_{kij}v^{j}Y^{k}+g_{0[k,i]}Y^{k}+O(5),
\end{eqnarray}
where $\tilde{\Gamma}_{ijk}=\frac{1}{2}(\tilde{\delta}_{ij,k}+\tilde{\delta}%
_{ik,j}-\tilde{\delta}_{jk,i})$. The magnitude of the 4-vector $Y^{\mu }$
should remain constant along the curve specified by $u^{\mu }$, so that 
\begin{equation}
\frac{d}{d\tau }\left( g^{\mu \nu }Y_{\mu }Y_{\nu }\right) =0,
\end{equation}%
or, up to $O(2)$, 
\begin{equation}
-(v^{i}Y_{i})^{2}+(1-2U)\tilde{\delta}^{ij}Y_{i}Y_{j}=\text{constant}.
\end{equation}%
This implies that a spin 3-vector, $S^{i}$, with constant magnitude, $\tilde{%
\delta}^{ij}S_{i}S_{j}$, is given by 
\begin{equation}
Y_{i}=(1+U)S_{i}+\frac{1}{2}\tilde{\delta}_{ij}v^{j}v^{k}S_{k}+O(4).
\end{equation}%
Alternatively, inverting this expression gives 
\begin{eqnarray}
S_{i} &=&(1-U)Y_{i}-\frac{1}{2}\tilde{\delta}_{ij}v^{j}v^{k}Y_{k}+O(4) 
\nonumber \\
&=&(1+U)\tilde{\delta}_{ij}Y^{j}-\frac{1}{2}\tilde{\delta}_{ij}\tilde{\delta}%
_{km}v^{j}v^{k}Y^{m}+O(4),
\end{eqnarray}%
which can now be differentiated with respect to $t$, and $dY_{i}/dt$ can be
substituted from above, to give 
\begin{equation}
\frac{dS_{i}}{dt}=\left[ 3v^{j}\tilde{\delta}^{km}\tilde{\delta}%
_{j[k}U_{,i]}+\tilde{\delta}^{km}g_{0[k,i]}+{\tilde{\Gamma}^{m}}_{ij}v^{j}%
\right] S_{m}+O(4).
\end{equation}%
where we have used 
\begin{equation}
\frac{dv^{i}}{dt}=\tilde{\delta}^{ij}U_{j}-{\tilde{\Gamma}^{i}}%
_{jk}v^{j}v^{k}.
\end{equation}%
For shorthand, one can also write 
\begin{equation}
\hat{\Omega}_{ki}\equiv 3v^{j}\tilde{\delta}_{j[k}U_{,i]}+g_{0[k,i]}
\end{equation}%
and 
\begin{equation}
\Omega ^{j}\equiv \frac{1}{2}\epsilon ^{jki}\hat{\Omega}_{ki}
\end{equation}%
so that 
\begin{equation}
\frac{dS_{i}}{dt}=\left[ \Omega ^{j}\epsilon _{jki}\tilde{\delta}^{km}+{%
\tilde{\Gamma}^{m}}_{ij}v^{j}\right] S_{m}+O(4).  \label{spin}
\end{equation}%
This is the general expression for spin precession in the geometry (\ref{geo}%
).

To see the effect of a deficit angle on spin precession we can now consider
a 3-metric that is locally isometric to Euclidean 3-space under the
coordinate redefinition (\ref{redef}). For an elliptic orbit in a plane of
constant $\theta$, and with a potential of the form $U=U(r)$, we
then have $v^{\theta }=0$ and $U_{,\theta }=U_{,\phi }=0$. From (\ref{spin}%
), we then find
\begin{eqnarray}
\frac{dS_{r}}{dt} &=&\frac{3}{2}v^{\phi }U_{,r}S_{\phi }+\frac{v^{\phi }}{r}%
S_{\phi } \\
\frac{dS_{\phi }}{dt} &=&-\frac{3}{2}v^{\phi
}r^{2}w^{2}U_{,r}S_{r}-rw^{2}v^{\phi }S_{r}+\frac{v^{r}}{r}S_{\phi }.
\end{eqnarray}%
If we now define two new quantities, $\tilde{S}_{\phi }\equiv S_{\phi }/(rw)$
and $X\equiv (3v^{\phi }rU_{,r}/2+v^{\phi })w,$ these equations reduce to 
\begin{eqnarray}
\frac{dS_{r}}{dt} &=&X\tilde{S}_{\phi } \\
\frac{d\tilde{S}_{\phi }}{dt} &=&-XS_{r},
\end{eqnarray}%
which have the solutions 
\begin{eqnarray}
S_{r} &=&A\sin \left\{ \int Xdt\right\} \\
\tilde{S}_{\phi } &=&A\cos \left\{ \int Xdt\right\}
\end{eqnarray}%
where $A$ is a constant of integration. If we now use (\ref{w}), with $%
\delta w$ being $O(2)$ small, then we find that the precession rate
integrated over an orbit is 
\begin{eqnarray}
\Omega &=& 2\pi -\int Xdt  \nonumber \\
&=& 2\pi -\frac{3}{2}\int v^{\phi }rU_{,r}dt-(1-\delta w)\int v^{\phi }dt 
\nonumber \\
&=& \Omega _{GR}+2\pi \delta w,
\end{eqnarray}%
where $\Omega _{GR}$ is the general relativistic precession rate, when $w=0$%
. The extra precession per orbit due to the deficit angle is therefore $2\pi
\delta w$.

This result can be seen to be straightforwardly generalisable to more
complicated orbits if we note that $w$ always enters as a multiplicative
factor in the $O(2)$ terms in (\ref{spin}). The lowest-order effect of $w$
then occurs when it multiplies the $O(0)$ terms, which are independent of
gravitational potentials and velocities.

\subsection{Orbital Inclination}

Binary pulsar observations are sensitive to the $sine$ of the inclination of
the orbital plane with respect to the line of sight of the observer: 
\begin{equation}
s=\sin i=\frac{x}{a},
\end{equation}%
where $a$ is the semi-major axis of the elliptical orbit, and $x$ is the
`projected' semi-major axis.  Although the value of this quantity
itself is not dependent on integration over an orbit, it is usually
determined in terms of another quantity that is: The orbital period.

In the Newtonian 2-body problem the orbital period is given by 
\begin{equation}
\label{Pb}
P_{b}=2\pi \sqrt{\frac{a^{3}}{m}},
\end{equation}%
where $m=m_{p}+m_{c}$ is the total mass of the pulsar and its companion
(another pulsar, in the case of the double pulsar). If there is a deficit
angle then we should expect this period to be reduced by a factor of $w$,
due to the shorter path length, so that 
\begin{equation}
P_{b}=2\pi w\sqrt{\frac{a^{3}}{m}}.
\end{equation}%
Using this to eliminate $a$ in (\ref{Pb}) we then find that 
\begin{equation}
s=w^{2/3}\left( \frac{2\pi }{P_{b}}\right) ^{2/3}x_{p}\frac{%
(m_{p}+m_{c})^{2/3}}{m_{c}},
\end{equation}%
where we have also used $x_{p}=xm_{c}/(m_{p}+m_{c})$, which is a directly
measurable quantity.


\subsection{Constraints}

The constraints on periastron advance, inclination angle and geodetic
precession rate obtained in \cite{1} and \cite{2} for the double pulsar
system are displayed in Table \ref{pulsres}. The first two of these are
based on observations made between April 2003 and January 2006 using the 64m
Parkes radio telescope in New South Wales, the 76m Lovell radio telescope at
the Jodrell Bank Observatory, and the 100m Green Bank telescope in West
Virginia. The observations of geodetic precession were made between December
2003 and November 2007 using the Green Bank telescope.

\begin{table}[tbh]
\begin{center}
\begin{tabular}{|l|l|}
\hline
\qquad Parameter & \qquad Observed value \\ \hline
Periastron advance, $\dot{\omega}$ & 16.89947 (68) $^{\circ}$/year \\ 
Orbital inclination, $s=\sin i$ & 0.99974 (+16,-39) \\ 
Geodetic precession rate, $\Omega_B$ & 4.77 (+66,-65) $^{\circ}$/year \\ 
\hline
\end{tabular}%
\end{center}
\caption{Observational constraints on the periastron advance, $\dot{\protect%
\omega}$, orbital inclination, $s=\sin i$, and geodetic precession rate of
PSR J$0737$-$3039$B, $\Omega _{B}$, from \protect\cite{1} and \protect\cite{2}%
. }
\label{pulsres}
\end{table}

Using the expressions found above, together with the usual general
relativistic ($\delta w = 0$) results, we can now write down the
periastron advance, inclination angle and geodetic precession rates as 
\begin{eqnarray}
\dot{\omega} &=&3\left( \frac{2\pi }{P_{b}}\right) ^{5/3}\frac{%
(m_{p}+m_{c})^{2/3}}{(1-e^{2})}+\left( \frac{2\pi }{P_{b}}\right) \delta w,
\\
s &=&\left( 1-\frac{2}{3}\delta w\right) \left( \frac{2\pi }{P_{b}}\right)
^{2/3}x_{p}\frac{(m_{p}+m_{c})^{2/3}}{m_{c}},
\end{eqnarray}%
and 
\begin{equation}
\Omega _{B}=\left( \frac{2\pi }{P_{b}}\right) ^{5/3}\frac{1}{(1-e^{2})}\frac{%
m_{c}(4m_{p}+3m_{c})}{2(m_{p}+m_{c})^{4/3}}+\left( \frac{2\pi }{P_{b}}%
\right) \delta w.
\end{equation}%
In all of these expressions $P_{b}$, $x_{B}$ and $e$ can be measured
directly, and are given in \cite{1}. The ratio of masses $m_{A}/m_{B}$ is
also known \cite{1}, but not the individual masses themselves. We therefore
need two observations to obtain a constraint on the deficit angle. The first
will give enough information to find the two masses, and the second can be
used to compare predicted and observed values of a relativistic effect. We
choose to use $\dot{\omega}$ as the first observable in each case, as it is
known to greatest accuracy.

Using the additional constraint from geodetic precession, $\Omega _{B}$,
then gives the $2\sigma $ bound \footnote{%
In order to combine asymmetric errors we have assumed they are Gaussian
distributed on each branch, with different variances.}
\begin{equation}
0\leq \delta w<1.5\times 10^{-6}.
\end{equation}%
Alternatively, we can use the inclination angle, $s$, to get the stronger $%
2\sigma $ bound 
\begin{equation}
0\leq \delta w<2.4\times 10^{-8}.
\end{equation}%

\vspace{5pt}
\section{Discussion}
\label{conc}

We have considered the effects of a deficit angle, $w$, in the space-time
metric for the solar system and the double pulsar system. Using observations
of the perihelion precession of Mercury and the gravitational deflection of
light we constrain the magnitude of such a deficit angle in the solar system
to be $2\pi (1-w)$, with 
\begin{equation}
0\leq (1-w)<10^{-9}
\end{equation}%
to $95\%$ confidence. Similarly, we have used observations of periastron
advance and inclination angle in the double pulsar system PSR J$0737$-$3039$%
A/B to gain the constraint 
\begin{equation}
0\leq (1-w)<2.4\times 10^{-8},
\end{equation}%
also to $95\%$ confidence. Although this result is weaker than the solar
system bound, it is in a very different physical environment, where accuracy
is likely to improve in the future, as observational data on the double
pulsar accumulates.

\section*{Acknowledgements}

We are grateful to Michael Kramer for helpful information. TC acknowledges
the support of Jesus College, Oxford and the BIPAC.

\end{document}